\documentclass[aps,pre,reprint]{revtex4-1}
\usepackage{amssymb}
\usepackage{amsmath}
\usepackage{amsthm}
\usepackage{nicefrac}
\usepackage{graphicx}
\usepackage{epstopdf}

\makeatletter
\newcommand*{\citenumns}[2][]{%
  \begingroup
  \let\NAT@mbox=\mbox
  \let\@cite\NAT@citenum
  \let\NAT@space\NAT@spacechar
  \let\NAT@super@kern\relax
  \renewcommand\NAT@open{}%
  \renewcommand\NAT@close{}%
  \cite[#1]{#2}%
  \endgroup
}
\makeatother

\begin{document}

\title{Compressing turbulence and sudden viscous dissipation with compression-dependent ionization state}

\author{Seth Davidovits}
\affiliation{Princeton University, Princeton, New Jersey 08544, USA}
\author{Nathaniel J. Fisch}
\affiliation{Princeton University, Princeton, New Jersey 08544, USA}
\affiliation{Princeton Plasma Physics Laboratory, Princeton, New Jersey 08544, USA}

\begin{abstract}
Turbulent plasma flow, amplified by rapid 3D compression, can be suddenly dissipated under continuing compression. This effect relies on the sensitivity of the plasma viscosity to the temperature, $\mu \sim T^{5/2}$. The plasma viscosity is also sensitive to the plasma ionization state. We show that the sudden dissipation phenomenon may be prevented when the plasma ionization state increases during compression, and demonstrate the regime of net viscosity dependence on compression where sudden dissipation is guaranteed. Additionally, it is shown that, compared to cases with no ionization, ionization during compression is associated with larger increases in turbulent energy, and can make the difference between growing and decreasing turbulent energy.
\end{abstract}

\maketitle

\section{Introduction}\label{sec:introduction}
Recently, simulations of compressing turbulent plasma demonstrated a sudden dissipation mechanism, which may enable a new paradigm for fast ignition inertial fusion~\cite{davidovits2016}. A plasma with initial (turbulent) flow is compressed on a timescale that is much faster than the dissipation time of the flow. This amplifies the turbulent kinetic energy (TKE) in the flow, for an ideal gas with subsonic flows. In a very rapid three dimensional adiabatic compression the energy in the flow scales at the same rate as the temperature. As the temperature increases, the plasma viscosity, which starts small, grows, because it scales as $\mu \sim T^{5/2}$. The viscosity first dissipates the smaller scales in the flow, which do not contain much energy. Eventually, as the compression continues, the energy-containing (largest) scales become viscous, and at this time all the TKE very suddenly dissipates into temperature. By initially putting most of the plasma energy in TKE, it may be possible to keep the plasma comparatively cool up until the sudden dissipation event, at which point it would ignite fusion or produce a burst of X-rays~\cite{davidovits2016}.

However, in addition to the temperature, the plasma charge state, $Z$, factors strongly into the viscosity, $\mu \sim T^{5/2}/Z^4$.
Laser and magnetically driven fusion experiments typically compress deuterium and tritium, with $Z=1$, so that, ignoring contaminants from the shell, the charge state is constant during the compression. In contrast, compression experiments designed to produce X-rays use a variety of higher $Z$ materials, which increase in ionization state during the compression. 

This increase in ionization state has the effect of slowing the viscosity growth. Consider, for example, a neon gas-puff Z-pinch~\cite{kroupp2011} that starts with $T \sim 13$ eV and $Z\sim3$, and finishes with $T \sim 200$ eV and $Z\sim9$. The temperature increase causes a growth in the viscosity by a factor of $\sim900$, while the (mean) ionization state growth reduces the viscosity by a factor of $\sim80$, drastically cutting the overall viscosity increase. 

In the present work, as in Ref.~[\citenumns{davidovits2016}], we consider a plasma temperature that increases due to the 3D adiabatic compression of an ideal gas in a box of side length $L$, going as $T = T_0 \left(L_0/L\right)^2$. The (mean) ionization state, $Z$, is treated as having some dependence on $L$ (i.e., the amount of compression) as well. This dependence is treated as fittable with some power, $Z=Z_0 \left(L_0/L\right)^{\zeta}$. Then, defining $\beta = \left(5 - 4\zeta \right)/2$, the viscosity can be written
\begin{equation}
\mu = \mu_0 \left(L_0/L\right)^{2 \beta} = \mu_0 \left(T/T_0\right)^{\beta}. \label{eq:mu}
\end{equation}
In this model, regarding the ionization state as a function of L is equivalent to regarding it as a function of T ($Z=Z_0 \left(T/T_0\right)^{\zeta/2}$), because $T\propto 1/L^2$.

Ionization processes in Z-pinch~\cite{foord1994} and laser driven plasmas are not simply temperature dependent, depending on density and more complex processes (e.g. shock dynamics). However, if the (mean) ionization state for a given experiment can be reasonably fit to $L$ as described, then the net effect in the present model is that the overall temperature dependence of the viscosity can be treated as some power other than 5/2. We expect $\beta \leq 5/2$, reflecting the assumption that the charge state increases under increasing compression (or temperature). For a rough estimate of a possible value for $\zeta$ and therefore $\beta$, consider that the first 26 ionization states of krypton (covering ~13 eV - 1200 eV) can be fit with $Z\sim T^{0.59}$. This corresponds to $\zeta = 1.18$, and $\beta = 0.14$. Since the ionization state can be higher at a given temperature than one would predict purely based on comparing the temperature to the ionization energies, one expects based on this example that a wide range of $\beta$ is possible in experiments, possibly including negative values.

Note that, if the adiabatic index of the compression is smaller than the value of $5/3$ assumed here, this also weakens the scaling of the viscosity with compression, effectively lowering $\beta$ ($\beta$ is defined so that $\mu \propto 1/L^{2\beta}$).

We consider initially turbulent plasma undergoing rapid, constant velocity, 3D isotropic compression and described by the same model as in Davidovits and Fisch~\cite{davidovits2016}, but with general $\beta$ rather than $\beta=5/2$. This model is described briefly in Sec. \ref{sec:model and energy}, and a derivation is given in the Appendix Sec. \ref{sec:derivation}. We show that there will be an eventual sudden dissipation when $\beta > 1$. For identical initial condition, starting viscosity, and compression velocity, lower $\beta$ cases show larger TKE growth and later sudden dissipation (when $\beta$ is still $>1$). Additionally, lower $\beta$ cases can show TKE growth under compression rates that would lead to the TKE damping in higher $\beta$ cases. For $\beta=1$, the TKE reaches a statistical steady state under constant velocity compression, for any compression rate above a threshold which we determine. When $\beta<1$, it seems there is no sudden dissipation, with the TKE increasing indefinitely instead. 

There are a number of implications of these results. The plasma in magnetically driven~\cite{kroupp2007,kroupp2007a,kroupp2011,maron2013} or laser driven~\cite{thomas2012,weber2014} compressions can be turbulent. There can be substantial reduction in viscosity growth due to increasing ionization state for a gas-puff Z-pinch. To the extent the turbulence generation mechanism(s) of a given compression approach is insensitive to $Z$, the present results show that, for a fixed rapid compression rate, a larger increase in $Z$ (weaker viscosity growth) is expected to correspond to larger TKE growth. Furthermore, increases in $Z$ can make the difference between growing or decaying TKE. 

Note that while these gas-puff Z-pinches appear to have substantial non-radial TKE even at stagnation~\cite{maron2013}, turbulence in the hot spot of ignition shots at the National Ignition Facility (NIF) is expected to be dissipated by high viscosity~\cite{weber2014}. The much higher temperatures in these hot spots create this high viscosity, but they are assisted by fuel of $Z=1$, to the extent it is not contaminated by mix. Our results demonstrate that even moderate reductions in the effective power $\beta$ from $5/2$ can cause large differences in TKE growth, and can determine whether for a given amount of compression ($L_{final}/L_0$) one can reach the dissipation regime.

The analysis in this work is carried out for 3D compressions, and as such is not strictly applicable to 2D compressions such as those in Z-pinches. In a 2D compression, the relative scaling of the TKE with compression, compared to the temperature is different (if the temperature growth is still assumed to be adiabatic and isotropic, the latter now being a larger assumption, since the plasma is driven anisotropically). Nevertheless, the intuition developed here may still be useful; all else being equal, more ionization enhances TKE growth under rapid compression by weakening the viscosity growth. The present work also neglects magnetic field effects, in line with other studies of turbulence in 3D compressions~\cite{thomas2012,weber2014}. This, too, limits the applicability to Z-pinch compressions, though there are also many instances in which the magnetic field need not dominate the dynamics in a Z-pinch~\cite{maron2013}.

The evolving ionization state during compression may be exploitable to optimize TKE growth before sudden dissipation, and to control the timing of the dissipation. If the ions in the compression become maximally ionized, then the viscosity change reverts to being dominated by temperature, while TKE growth up to this point will be larger than without ionization. Mixes of ion species open up a wide range of control possibilities for the viscosity dependence on compression, but also introduce other complications (e.g. species separation), and are beyond the scope of the present work. However, the prospect of controlling ion charge state and thereby viscosity appears to enlarge considerably the parameter space of opportunities for optimizing both the energy and pulse length in a sudden dissipation resulting in X-ray emission.

The structure of the paper is as follows. Section \ref{sec:model and energy} gives a brief description of the model, and discusses the energy equation for the turbulence, which is used in Sec. \ref{sec:analysis} to show some analytic results and to describe the general phenomenology. To go along with this analysis, the results from numerical simulations of compressing turbulence with ionization are displayed in Figs.~\ref{fig:KEvsL_twoBetas}, \ref{fig:KEvsL_beta1} and \ref{fig:vary_beta} and discussed in the captions and Sec. \ref{sec:simulations}. Section \ref{sec:discussion} discusses implications of the results and caveats associated with them. Some secondary calculations associated with Secs.~\ref{sec:model and energy} and \ref{sec:analysis} are contained in the Appendix, and referenced at the appropriate point.

\section{Model and energy equation}\label{sec:model and energy}
\subsection{Model}
The model used here follows previous work by Wu~\cite{wu1985} and others~\cite{coleman1991,blaisdell1991,cambon1992,hamlington2014}, and is the same as that in Davidovits and Fisch~\cite{davidovits2016}, allowing for a general power $\beta$ for the viscosity dependence on temperature. For completeness a derivation is given in the Appendix section \ref{sec:derivation}.

The essence of the model is as follows. It describes the 3D, isotropic compression of homogeneous turbulence in the limit where the turbulence Mach number goes to zero. Compression is achieved through an imposed background flowfield. The effect of the flow is that a cube, of initial side length $L_0$, will shrink in time but remain a cube. The side length of the box as a function of time will be
\begin{equation}
L\!\left(t\right) = L_0 - 2 U_b t,
\end{equation}
where $U_b$ is the (constant) velocity of each side of the cube. In the low Mach limit, density fluctuations are ignored, and the density increases in time as one would expect for the compression,
\begin{equation}
\rho_0\!\left(t\right) = \rho_0\!\left(0\right) \left(L_0/L\!\left(t\right)\right)^3.\label{eq:mean_density_solution}
\end{equation}
The temperature of the compressing plasma is that for adiabatic compression of an ideal gas,
\begin{equation}
T\!\left(t\right) = T_0 \left(L_0/L\left(t\right)\right)^2.\label{eq:temperature_solution}
\end{equation}
The viscosity dependence on $L$ (alternatively, $T$) is given by Eq. (\ref{eq:mu}).

The evolution of the initially turbulent flow is solved in a frame that moves along with the background flow, on a domain that extends from $\left[-L_0/2,L_0/2\right]$ in each dimension and has periodic boundary conditions. The energy in the turbulence in this frame is the same as in the lab frame. In this frame, after using Eqs.~(\ref{eq:mu},\ref{eq:mean_density_solution},\ref{eq:temperature_solution}) to write the density, temperature and viscosity dependence in terms of $L$, the Navier-Stokes equation for the turbulence is
\begin{equation}
\frac{\partial \mathbf{V}}{\partial t}+\frac{1}{\bar{L}}\mathbf{V}\cdot \nabla \mathbf{V}-\frac{2U_{b}}{L}\mathbf{V}+\frac{\bar{L}^2}{\rho_{0}\!\left(0\right)} \nabla P  =  \nu_0 \left(\frac{1}{\bar{L}}\right)^{2\beta-1}\nabla^2 \mathbf{V}.\label{eq:moving_momentum}
\end{equation}
The initial kinematic viscosity is $\nu_0=\mu_0/\rho_0\!\left(0\right)$, and $\bar{L}=L/L_0$.

\subsection{Energy equation}
The energy density in the fluctuating flow, calculated in the moving
frame is $E=\rho_{0}\!\left(0\right)\mathbf{V}^{2}/2$. The total
energy is then $E^{T}=\iiint_{-L_{0}/2}^{L_{0}/2}\mbox{d}\mathbf{x}E$.
Since $\mathbf{v}'=\mathbf{V}$ (see Appendix Sec. \ref{sec:derivation}), this total energy is the same as
the total energy in the lab frame (in the lab frame, the density increases,
but the volume to be integrated decreases in a manner that balances
it). The time evolution of the energy density is
\begin{equation}
\frac{\partial E}{\partial t}=\rho_{0}\!\left(0\right)\mathbf{V}\cdot\frac{\partial\mathbf{V}}{\partial t}.
\end{equation}
Equation (\ref{eq:moving_momentum}) is used to write this energy equation explicitly. In Fourier ($k$, wavenumber) space, since the flow is assumed to be homogeneous and isotropic, it is
\begin{equation}
\frac{\partial E\!\left(k,t\right)}{\partial t}=\frac{T\!\left(k,t\right)}{\bar{L}}+\frac{4U_{b}}{L}E\!\left(k,t\right)-2\nu_{0}\bar{L}^{1-2\beta}k^{2}E\!\left(k,t\right),
\end{equation}
with $T\!\left(k,t\right)$ a nonlinear term that includes the effects of the pressure and $\mathbf{V}\cdot\nabla\mathbf{V}$
terms (see, e.g. McComb~\cite{mccomb1990}). The effect of $T\!\left(k,t\right)$ is to transfer energy
between wavenumbers (modes), conservatively. Integrated over the whole
of $k$ space, it vanishes. The total energy is
\begin{equation}
E^{T}\!\left(t\right) = \int_{k_{\mathrm{min}}}^{\infty}\mbox{d}kE\!\left(k,t\right).
\end{equation}
In the moving frame, $k_{\mathrm{min}}=2\pi/L_{0}$ is fixed, given by the initial
size of compressing system, e.g. capsule (although the current model uses periodic boundaries). In principle structures can be arbitrary small, so $k_{\mathrm{max}}=\infty$, but practically $E\!\left(k,t\right)$ will be zero above some $k$. The evolution of the total energy is,
\begin{equation}
\frac{\mbox{d}E^{T}\!\left(t\right)}{\mbox{d}t} = \int_{k_{\mathrm{min}}}^{\infty}\mbox{d}k\left(\frac{4U_{b}}{L}-2\nu_{0}\bar{L}^{1-2\beta}k^{2}\right)E\!\left(k,t\right).\label{eq:integrated_energy}
\end{equation} 

\begin{figure*}
\includegraphics[]{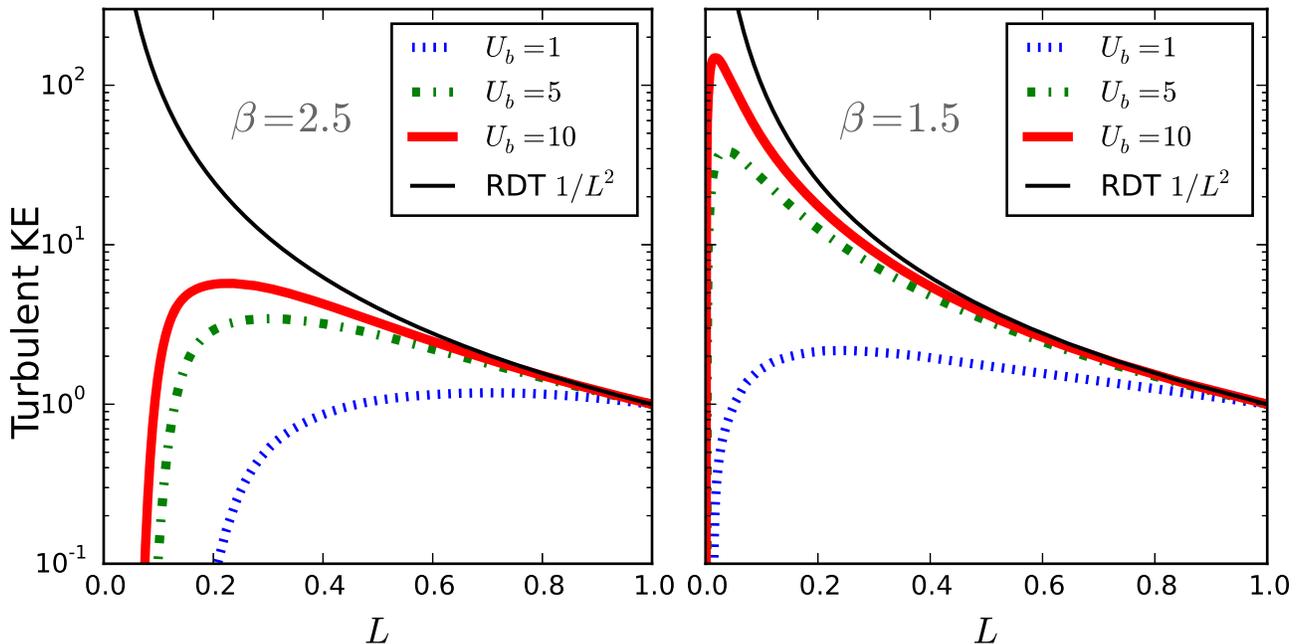}
\caption{Turbulent kinetic energy (TKE) during compression at various rates, for two different effective viscosity dependencies on temperature, $\mu\!\left(T\right) \sim T^\beta$, representing ionization effects (see Eq.~(\ref{eq:mu}) and the surrounding discussion). On the left, $\beta=2.5$, the plasma case with no ionization effects. On the right, $\beta =1.5$. An initial flow field, with TKE normalized to 1, is compressed with velocity $U_b$ on times equal to ($U_b=1$) and faster than ($U_b = 5,10$) the initial turbulence decay time (compression times $L_0/\left(2U_b\right)$ are normalized to the initial turbulent decay time). The initial domain is a box of size $L_0^3=1^3$, time progresses right to left ($t=(1-L)/(2U_b)$) as the compression shrinks the domain. The same initial flow field is used for all compressions, so that the only difference is $\beta$. All cases show an eventual sudden dissipation of the TKE. For a given compression velocity, the lower $\beta$ case shows stronger TKE growth and a later, more sudden, dissipation. For this initial Reynolds number (600), the change from $\beta=2.5$ to $\beta=1.5$ pushes the dissipation $L$ for $U_b=5,10$ from $\sim0.1$ to $<0.01$, which could make the difference between dissipating during a compression or not. Similarly, for $U_b=1$, the change from $\beta=2.5$ to $\beta=1.5$ greatly increases the compression needed to reach the point where the TKE dissipates.
The theoretical rapid distortion theory~\cite{durbin2010,savill1987,hunt1990} (RDT) solution is shown for comparison (in this case, it is the solution to Eq.~(\ref{eq:integrated_energy}), neglecting the dissipation, see also Wu~\cite{wu1985}). It gives the theoretical maximum growth of the TKE with the compression.
\label{fig:KEvsL_twoBetas}}
\end{figure*}

\begin{figure}
\includegraphics[]{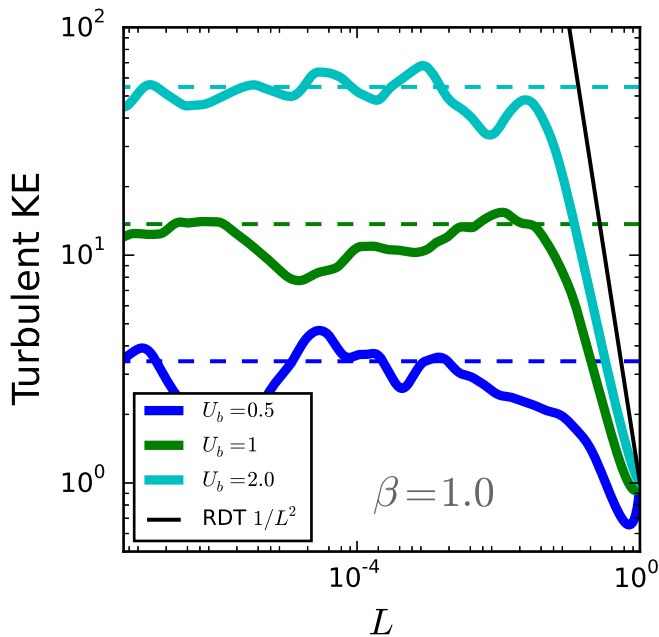}
\caption{Same as Fig.~\ref{fig:KEvsL_twoBetas}, but for $\beta=1.0$, at a lower Reynolds number, and with a logarithmic scale for $L$. No eventual sudden dissipation is observed, even after extreme amounts of compression. After an initial growth phase, the turbulent kinetic energy (TKE) saturates and fluctuates around the mean level predicted by Eq.~(\ref{eq:E_steady_Ub}).  This theoretically predicted mean level of the TKE is shown as a dotted line for each compression velocity. Note Eq.~(\ref{eq:E_steady_Ub}) must be written in the same velocity normalization as the figure before being applied.
\label{fig:KEvsL_beta1}}
\end{figure}

\begin{figure}
\includegraphics[]{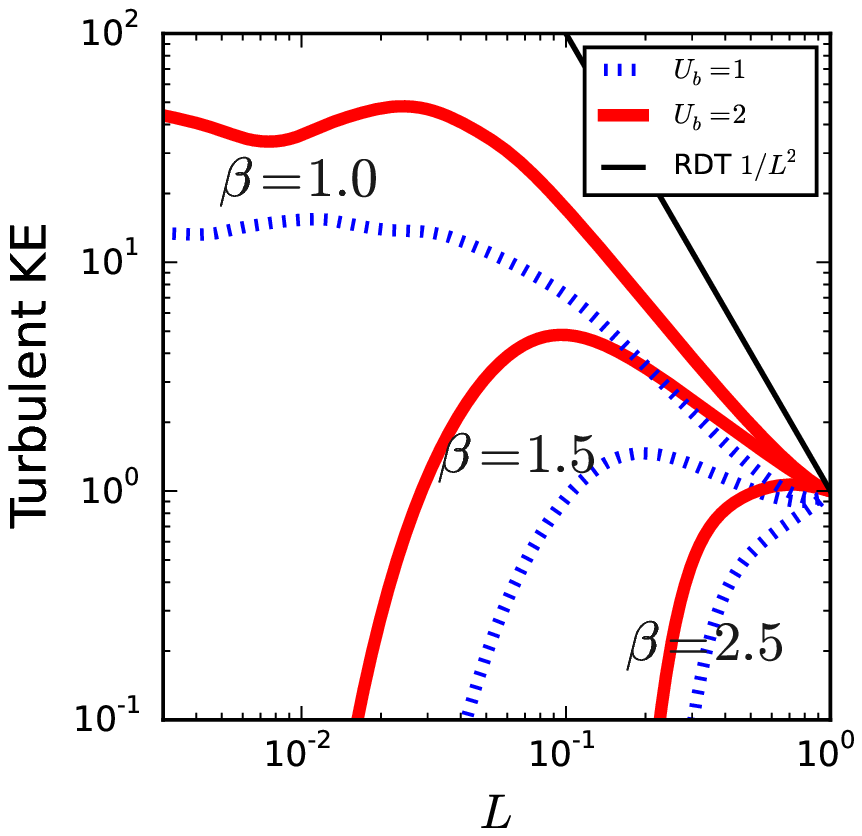}
\caption{Similar to Figs.~\ref{fig:KEvsL_twoBetas},~\ref{fig:KEvsL_beta1}; the turbulent kinetic energy (TKE) for the same initial condition compressed at two different rates and different values of $\beta$, showing the effect of varying the amount of ionization during compression ($\beta$). The red, solid lines use a compression time that is half the initial turbulent decay time, while the blue, dashed lines use a compression time that is the same as the initial turbulent decay time. For a given compression rate ($U_b=1$ or $U_b=2$), the TKE is larger at every stage of the compression when $\beta$ is lower (when there is more ionization during compression). For the case when $U_b = 1$, the TKE purely decays when $\beta = 2.5$ (the plasma case with no ionization). Ionization during compression can cause this to no longer be the case; when $\beta$ decreases to 1.5 or 1.0, the TKE either grows before dissipating, or grows without dissipating.
\label{fig:vary_beta}}
\end{figure}

\section{Analysis}\label{sec:analysis}
Since $E\!\left(k,t\right)\geq0$, the energy is guaranteed to decrease
if the coefficient of $E\!\left(k,t\right)$ in Eq. (\ref{eq:integrated_energy}) is negative for all $k\in\left[k_{\mathrm{min}},\infty\right]$;
conversely, it is guaranteed to increase if the coefficient is
positive for all $k$ where $E\!\left(k,t\right) \neq 0$. (However, this latter
condition is difficult to work with, since for $k\rightarrow\infty$
there is always damping and as the energy increases $T\!\left(k,t\right)$ will tend to move energy to higher $k$). These conditions are sufficient, but not necessary. The guaranteed decrease condition requires that for every mode
\begin{equation}
\frac{2U_{b}/L_{0}}{\nu_{0}k^{2}} < \bar{L}^{2-2\beta}.\label{eq:decrease_condition}
\end{equation}
The left hand side is largest for $k=k_{\mathrm{min}}$, and trends to 0 as $k\rightarrow\infty$. 

\subsection{$\beta > 1$}
When $\beta>1$, the right hand side of Eq. (\ref{eq:decrease_condition}) starts at 1 at $t=0$ and increases towards $\infty$ as $L\rightarrow0$. At some time the condition will be satisfied for all $k$, when
\begin{equation}
\frac{2U_{b}/L_{0}}{\nu_{0}k_{\mathrm{min}}^{2}}=\frac{1}{\bar{L}^{2\beta-2}}
\end{equation}
Thus, the energy will always decay eventually for fixed $\nu_{0},U_{b},k_{\mathrm{min}}$
when $\beta>1$. This is not to say that the energy cannot decrease
before this condition is satisfied.

\subsection{$\beta = 1$}\label{sec:analysis_beta1}
When $\beta=1$, the right hand side of Eq.~(\ref{eq:decrease_condition}) is 1. In this case there is no time dependence in the condition for guaranteed energy decrease. If Eq.~(\ref{eq:decrease_condition}) is initially satisfied for all $k$, the energy will purely decay, with no initial growth phase. 

Otherwise, a fixed range of wavenumbers have a net positive coefficient for $E\!\left(k,t\right)$ in Eq.~(\ref{eq:integrated_energy}), while the rest have a net negative coefficient (ignoring the nonlinearity). The wavenumber cutoff between these two regions is given by equality in Eq.~(\ref{eq:decrease_condition}),
\begin{equation}
k_{\mathrm{cutoff}}=\left(\frac{4U_{b}/L_{0}}{2\nu_{0}}\right)^{1/2}. \label{eq:k_cutoff}
\end{equation}
The width of wavenumbers with a net forcing (linearly) is $\Delta k_{\mathrm{forced}}=k_{\mathrm{cutoff}}-k_{\mathrm{min}}$.
Since the range of net forced wavenumbers is fixed, it might be expected that the energy will reach a (statistical) steady state. This is the case, and it can be shown (see Appendix Sec. \ref{sec:beta_1_steady}) that the statistically steady state energy is
\begin{equation}
E^T_{\mathrm{steady}} = \left(1.9\right) \rho_0 \! \left( 0 \right) U_b^2. \label{eq:E_steady_Ub}
\end{equation}
Further, the spectrum itself, $E\!\left(k,t\right)$, converges to a statistical steady state $E\!\left(k\right)$ in simulations~\cite{rosales2005}. While the steady state energy is independent of the viscosity $\nu_0$ (alternatively, the initial Reynolds number), the details of the energy spectrum of the saturated turbulence will not be. Also, as already mentioned, if the initial viscosity is too large, there is no steady state and the energy will purely decay.

This steady state energy can be rewritten in terms of $\Delta k_{\mathrm{forced}}$ by using Eq.~(\ref{eq:k_cutoff}),
\begin{equation}
E^{T}_{\mathrm{steady}} = \frac{\left(1.9\right)}{4}\rho_{0}\!\left(0\right)\nu_{0}^{2}L_{0}^{2}\left(\Delta k_{forced}+\frac{2\pi}{L_{0}}\right)^{4}
\end{equation}
Once the sign of the coefficient of $E\!\left(k,t\right)$ at a given $k$ in Eq.~(\ref{eq:integrated_energy}) is being considered (rather than the sign of \emph{all} coefficients), the nonlinearity cannot be ignored. Thus, $k_{\mathrm{cutoff}}$ is not necessarily a true (statistically steady state) cutoff between net forced and damped modes, but rather the linear cutoff.

\subsection{$\beta < 1$}
When $\beta<1$, the right hand side of Eq.~(\ref{eq:decrease_condition}) trends to 0 as time increases, and an increasing number of shorter wavelength modes will have a net forcing (ignoring the nonlinearity). This means that $\Delta k_{\mathrm{forced}}$ trends to infinity as $L\rightarrow0$. With the rather large caveats that in this case the problem is not an equilibrium one, and that the nonlinearity has been ignored in looking at the number of modes with a net forcing, the result from the equilibrium case that the steady state energy is proportional to the number of linearly forced modes suggests that the energy for $\beta<1$ continually increases for late times (after any initial transients are erased) under constant compression.

Note that neutral gas, compared to plasma with no ionization, has a weak dependence of viscosity on temperature, with studies of compressing gas turbulence using values that fall in the $\beta<1$ case (e.g. $\beta=3/4$~\cite{wu1985,coleman1991,coleman1993}). Turbulence closure models in these works, which include the evolution of the TKE in a neutral gas under compression, give a continually increasing TKE when evaluated for an initially rapid, constant velocity, 3D compression, consistent the suggestion here.

\section{Simulations} \label{sec:simulations}
In Section \ref{sec:analysis} we showed that; for $\beta >1$, the TKE should always eventually damp, even with continued constant velocity compression (which represents an ever increasing compressive force); and when $\beta = 1$, the TKE will either purely decay or reach a steady state under continued compression. We also suggested that the energy always increases under continued constant velocity compression when $\beta < 1$ (if the compression is initially rapid, if not, the energy may decrease for some period before eventually increasing). These represent different regimes of the viscosity dependence on compression - with little to no ionization during compression, $\beta$ will be near the ionization free value, $\beta = 5/2$, and the sudden viscous dissipation phenomenon will still be possible. If substantial ionization occurs during a phase of the compression, then $\beta$ may be significantly reduced from $5/2$, and the viscous dissipation of the TKE will be prevented.

In order to get a better sense of the effect of decreasing $\beta$, we perform direct numerical simulation of compressing turbulence for a few values of $\beta$. The scaled form of the momentum equation, Eq.~(\ref{eq:scaled_momentum}), is simulated with periodic boundary conditions using the spectral code Dedalus~\cite{dedalus}. Results are then translated back into the lab frame using the appropriate rescaling. Initial conditions are generated using the forcing method of Lundgren~\cite{lundgren2003,rosales2005}. All simulations are carried out on a $192^3$ Fourier grid, which is dealiased to $128^3$.

Simulations are done for three different values of $\beta$, $5/2$, $3/2$, and $1$. Of note is that for $\beta = 3/2$, the forcing term drops out of Eq.~(\ref{eq:scaled_momentum}), and it is simply the usual Navier-Stokes equation. This means that a single decaying turbulence simulation can give results for all compression velocities (at one initial Reynolds number).

Figures~\ref{fig:KEvsL_twoBetas},~\ref{fig:KEvsL_beta1} and~\ref{fig:vary_beta} and their captions describe results from these simulations. The simulations in Fig. \ref{fig:KEvsL_twoBetas} are carried out with an initial Reynolds number of 600. Those for Fig.~\ref{fig:KEvsL_beta1} are carried out with an initial Reynolds number of 100, which is necessary so that the turbulence remains fully resolved at saturation. Those in Fig. \ref{fig:vary_beta} also use an initial Reynolds number of 100, again to keep the $\beta=1.0$ case fully resolved at saturation.

\section{Discussion} \label{sec:discussion}
The present model ignores many effects that do or may play an important role during compression of plasmas. The suggested $\beta=1$ cutoff between eventually dissipating and perpetually growing TKE need not hold true in a more complete model. Non-ideal equation of state effects are neglected. It should be noted that only constant velocity compressions were considered; compressions with time-dependent velocities would also change the cutoff. Boundary effects, which are ignored, would be expected to become increasingly important as the amount of compression increased.  The manner in which the ionization is accounted for neglects, among other effects, the energy required to achieve the ionization. If this energy comes at the expense of the temperature, and the true rate of temperature increase is less than $\sim 1/L^2$, this would alter $\beta$, but the general idea remains the same. 

Subsonic compressions have been assumed, which is not necessarily the case for current compression experiments, nor is it the regime in which schemes utilizing the sudden dissipation effect would likely be operated. Because the compressions are subsonic, the feedback of the dissipated TKE into the temperature is also neglected, which is expected to only make the sudden dissipation, once it happens, even more sudden.

As previously discussed, magnetic effects have also been neglected. Although this may be reasonable for 3D compressions, and certain regimes of 2D Z-pinch compressions, to expand the study for general 2D compression (Z-pinch) applicability will require the inclusion of magnetic effects. The inclusion of magnetic fields through a magnetohydrodynamic (MHD) model will introduce a number of new considerations. If there is a strong background magnetic field, the plasma conditions can be highly anisotropic, and the intuition from the present discussion may be difficult to apply. With magnetic fields included, turbulent energy can be stored in fluctuations of the magnetic field, and turbulent dissipation can occur through the plasma diffusivity, $\eta$. Continuing to assume the incompressible limit, then, in addition to the Reynolds number, the magnetic Reynolds number $\mathrm{Re}_{m} = U \mathcal{L}/\eta$ and the magnetic Prandtl number $\mathrm{Pr}_m = \mathrm{Re}_m/\mathrm{Re} = \nu/\eta$ are also important for characterizing any turbulence.

The plasma magnetic diffusivity scales with ion charge state and plasma temperature as $\eta_{Spitzer} \sim Z/T^{3/2}$. Then, the magnetic Prandtl number has a temperature and charge state scaling of $\mathrm{Pr}_m \sim T^4/Z^5$. The behavior of MHD turbulence is influenced by the relative values of $\mathrm{Re}$, $\mathrm{Re}_m$, and $\mathrm{Pr}_m$; they affect whether magnetic fluctuations will grow (see e.g.~\cite{schekochihin2007,iskakov2007}), the saturated ratio of turbulent magnetic energy compared to kinetic energy, and the steady state ratio of viscous dissipation to dissipation through the magnetic diffusivity (see e.g.~\cite{brandenburg2014}). From the considerations in the present work, it is clear that, depending on the amount of ionization during compression, a range of behaviors for the dimensionless quantities is possible. Considering the limit of no ionization, and assuming the scalings $T\sim 1/L^2, \mathcal{L} \sim L, U \sim 1/L$, one has that: the viscosity increases with compression (and the Reynolds number decreases), the magnetic diffusivity decreases (and the magnetic Reynolds number increases), and the magnetic Prandtl number increases. At high magnetic Prandtl number and large magnetic Reynolds number (but assuming the Reynolds number is still large enough for turbulent flow), the small scale dynamo is effective, so that magnetic perturbations can grow up quickly and saturate, while the ratio of kinetic dissipation to magnetic dissipation appears to grow large~\cite{brandenburg2014}. Investigations of these effects are a subject of current research and debate, and are typically carried out in steady state, whereas the sudden dissipation effect relies on dynamics far from steady state. As such, specific investigations of MHD effects on sudden dissipation are needed before making predictions.

A simple model of the impact of radiation, the effects of which have been neglected in the preceding discussion, is included in the Appendix Sec.~\ref{sec:radiation}. This model consists of a temperature equation that includes mechanical heating and radiative cooling due to optically thin electron bremsstrahlung. With no radiation, the mechanical heating gives the $T\sim 1/L^2$ adiabatic temperature scaling. When the bremsstrahlung is included, it is shown that the temperature can still track closely with the adiabatic result for a large amount of the compression, provided the initial ratio of the radiation term to the mechanical heating term is very small. Then, the results in the present work will not be significantly modified. The ratio of radiation term to mechanical heating term, $\mathcal{R}$, can be written as $\mathcal{R} \sim (\tau_c/3.01\times 10^{-9}) (\rho_{g/cc}/A_i T_{keV}^{1/2}) (2 Z^3/(1 + Z))$. For details, see Appendix Sec.~\ref{sec:radiation}. Here $\tau_c$ is the compression time, $\rho$ the density, $T$ the temperature, $A_i$ the ion mass number, and $Z$ the ion charge state. These considerations are a subset of the usual power balance requirements for inertial confinement experiments (see, for example, Lindl~\cite{lindl1995}). In both cases, it is desirable to operate in parameter regimes where the temperature increases under compression when radiation effects are included. From the perspective of radiation, the presence and quantity of the hydrodynamic motion does not modify the potential operating regimes as compared to compression schemes without hydrodynamic motion. The same will be true with the inclusion of line-radiation, important for high-Z plasmas. 

However, the operating regimes where the temperature increases under compression will be modified by the turbulence in (at least) two ways that are neglected in this work. First, before any sudden dissipation event, there will be some level of viscous dissipation of hydrodynamic motion into temperature. When the hydrodynamic energy is large compared to the thermal energy, this heating may somewhat relax the operating regimes where the temperature increases under compression. On the other hand, a second neglected effect, turbulent heat transport, represents a cooling effect that opposes this heating effect.

Once the sudden dissipation event is triggered, in the supersonic case with the feedback of dissipated TKE into temperature included, the temperature should rise rapidly. Taking into account radiation will then be important for modeling the sudden dissipation event itself, which occurs over a small time interval so that the plasma volume hardly changes.

Despite these deficiencies, the present work serves to highlight the sensitivity of TKE growth under compression to changes in the viscosity scaling with compression, in which ionization can play a strong role. Thus, even modest contamination of a low-Z plasma with higher-Z constituents may have substantial hydrodynamic implications, as, say, atomic mix in an ICF hotspot. Finally, this sensitivity to the ionization state suggests that the possibilities for control of TKE growth and sudden dissipation for X-ray production are now significantly expanded. This expansion of possibilities comes in part from the prospect of considering a wide range of ion-species mixes. Although outside the scope of the present work, it can be anticipated that using a variety of mixtures could enable detailed and controlled shaping of the X-ray emission pulse.

\begin{acknowledgments}
This work was supported by DOE through Contracts No. DE-AC02-09CH1-1466 and NNSA 67350-9960 (Prime $\#$ DOE DE-NA0001836), by DTRA HDTRA1-11-1-0037, and by NSF Contract No. PHY-1506122.
\end{acknowledgments}

\appendix*
\section{Derivations}
\subsection{Model derivation}\label{sec:derivation}
Although essentially identical models have been discussed elsewhere~\cite{wu1985,coleman1991,blaisdell1991,cambon1992,hamlington2014}, for the sake of completeness, and to explain some details, we present a derivation here. Start with the continuity and momentum equations for compressible Navier-Stokes,
\begin{eqnarray}
\frac{\partial}{\partial t}\left(\rho\right)+\frac{\partial}{\partial x_{i}}\left(\rho v_{i}\right) & = & 0,\\
\frac{\partial}{\partial t}\left(\rho v_{i}\right)+\frac{\partial}{\partial x_{j}}\left(\rho v_{i}v_{j}\right) + \frac{\partial}{\partial x_{j}}\left(\delta_{ij}p\right) & = & D_{i}\\
\frac{\partial}{\partial x_{j}}\left(\mu\left(\frac{\partial v_{i}}{\partial x_{j}}+\frac{\partial v_{j}}{\partial x_{i}}-\frac{2}{3}\delta_{ij}\frac{\partial v_{k}}{\partial x_{k}}\right)\right) & = & D_{i}.
\end{eqnarray}
The Stokes' hypothesis has been used, that the second viscosity coefficient, often denoted $\lambda$, is $\lambda=-\frac{2}{3}\mu$. This form of the rate of strain tensor is consistent with the Braginskii result~\cite{braginskii1965}. The unknowns are rewritten as two parts,
\begin{eqnarray}
v_{i}\!\left(\mathbf{x},t\right) & = & v_{i0}\!\left(\mathbf{x},t\right)+v_{i}'\!\left(\mathbf{x},t\right)\\
\rho\!\left(\mathbf{x},t\right) & = & \rho_{0}\!\left(\mathbf{x},t\right)+\rho'\!\left(\mathbf{x},t\right)\\
p\!\left(\mathbf{x},t\right) & = & p_{0}\!\left(\mathbf{x},t\right)+p'\!\left(\mathbf{x},t\right)
\end{eqnarray}
where $v_{i0}$ is given, and the subscript 0 indicates ensemble averaged
quantities, while prime quantities have 0 ensemble average. The prime
quantities are assumed to be statistically homogeneous, and ultimately
the equations governing their evolution will have no explicit spatial
dependence, allowing the use of periodic boundary conditions. For
the prime quantities to be homogeneous, it can be shown (see, e.g. Blaisdell~\cite{blaisdell1991})
that the flow $v_{i0}$ must be of the form,
\begin{equation}
v_{i0}\!\left(\mathbf{x},t\right)=A_{ij}\!\left(t\right)x_{j}.
\end{equation}
For this work, only pure (no shear), isotropic compressions are considered, so that
\begin{equation}
A_{ij}\!\left(t\right)=a\!\left(t\right)\delta_{ij}
\end{equation}
with $\delta_{ij}$ the Kronecker delta. When $a\!\left(t\right)<0$ this enforced, ``background'', flow is compressive.

With these assumptions, the continuity equation is
\begin{equation}
\frac{\partial}{\partial t}\left(\rho_{0}+\rho'\right)+\frac{\partial}{\partial x_{i}}\left(\left(\rho_{0}+\rho'\right)a\!\left(t\right)x_{i}+\rho_{0}v_{i}'+\rho'v_{i}'\right) = 0.\label{eq:continuity}
\end{equation}
Taking an ensemble average gives an equation for $\rho_{0}$. Denoting
the average as $\langle\rangle$, then by definition $\langle\rho'\rangle=\langle v_{i}'\rangle=0$.
Also, $\partial\langle\rho'v_{i}'\rangle/\partial x_{i}=0$ because
ensemble averages, such as $\langle\rho'v_{i}'\rangle$, are assumed
to be homogeneous. The equation for $\rho_{0}$ is then,
\begin{equation}
\frac{\partial\rho_{0}}{\partial t}+a\!\left(t\right)x_{i}\frac{\partial\rho_{0}}{\partial x_{i}}+3a\!\left(t\right)\rho_{0} = 0.\label{eq:mean_continuity}
\end{equation}
It can be shown~\cite{blaisdell1991} that only for $\rho_{0}\!\left(\mathbf{x},t\right)=\rho_{0}\!\left(t\right)$
can the homogeneous turbulence constraint be satisfied. Dropping the second term in Eq.~(\ref{eq:mean_continuity}) accordingly, the density is
\begin{equation}
\rho_{0}\!\left(\mathbf{x},t\right)=\rho_{0}\!\left(t\right)=\rho_{0}\!\left(0\right)\exp\left[-3\int_{0}^{t}a\!\left(t'\right)\mbox{d}t'\right].\label{eq:mean_density_exp}
\end{equation}

The fluctuating density is determined by Eq.~(\ref{eq:continuity}), which can be
simplified by canceling the terms that sum to 0 according to Eq.~(\ref{eq:mean_continuity}).
It is
\begin{equation}
\frac{\partial\rho'}{\partial t}+\frac{\partial}{\partial x_{i}}\left(\rho'a\!\left(t\right)x_{i}+\rho_{0}v_{i}'+\rho'v_{i}'\right)=0.
\end{equation}
For incompressible fluctuating (non-background) flow, we assume that
the flow $v_{i}$ is low Mach, so that sound waves can be neglected
and the density perturbation $\rho'$ can be ignored. Then, the fluctuating
continuity equation reduces to the divergence free constraint on the
prime velocity,
\begin{equation}
\rho_{0}\frac{\partial}{\partial x_{i}}\left(v_{i}'\right)=0.
\end{equation}
With $v_{i0}$ as given, $\rho'\rightarrow0$, and $\rho_{0}$ depending only on time, the momentum equation is
\begin{multline}
\rho_{0}\left(\frac{\partial v_{i}'}{\partial t}+v_{j}'\frac{\partial v_{i}'}{\partial x_{j}}+\left(a^{2}+\dot{a}\right)x_{i}+ax_{j}\frac{\partial v_{i}'}{\partial x_{j}}+av_{i}'\right)+ \\
\frac{\partial}{\partial x_{j}}\left(\delta_{ij}\left(p_{0}+p'\right)\right) = D_{i}.\label{eq:momentum}
\end{multline}
The ensemble averaged momentum equation is
\begin{equation}
\rho_{0}\left(a^{2}+\dot{a}\right)x_{i}+\frac{\partial}{\partial x_{j}}\left(\delta_{ij}p_{0}\right) = 0.\label{eq:mean_momentum}
\end{equation}
In arriving at $\langle D_{i}\rangle=0$, the viscosity $\mu$ is assumed to be independent of space. The mean momentum equation, Eq.~(\ref{eq:mean_momentum}), says $p_{0}$ is quadratic in $\mathbf{x}$, unless
$a^{2}+\dot{a}=0$, in which case $p_{0}$ is independent of $\mathbf{x}$.
Since $a$ sets the time dependence of the background flow (the rate
of compression), this means only for one particular background flow
can $p_{0}$ be independent of $\mathbf{x}$. For the purposes of this work, we consider temperature dependent viscosity, $\mu=\mu\!\left(T\right)$.
The equation of state relates the pressure, density and temperature, $p=\rho RT$. This is $p_{0}+p'=\rho_{0}RT
$, which becomes, after taking the ensemble average,
\begin{equation}
p_{0}=\rho_{0}\!\left(t\right)R\langle T\rangle. \label{eq:mean_EOS}
\end{equation}
In order to have $T=T\!\left(t\right)$, so that $\mu\!\left(T\right)$ is independent of space, we must take
\begin{equation}
a^{2}+\dot{a}=0, \label{eq:flow_condition}
\end{equation}
so that $p_{0}=p_{0}\!\left(t\right)$. Then, Eqs.~(\ref{eq:mean_density_exp},\ref{eq:mean_EOS},\ref{eq:flow_condition}) and the condition for an adiabatic compression together determine $T$ and $p_{0}$.

Subtracting Eq.~(\ref{eq:mean_momentum}) from Eq.~(\ref{eq:momentum}) gives the equation governing the fluctuating flow,
\begin{multline}
\rho_{0}\left(\frac{\partial v_{i}'}{\partial t}+v_{j}'\frac{\partial v_{i}'}{\partial x_{j}}+ax_{j}\frac{\partial v_{i}'}{\partial x_{j}}+av_{i}'\right) = \\
- \frac{\partial}{\partial x_{j}}\left(\delta_{ij}p'\right)+ \mu\!\left(T\right)\frac{\partial^{2}v_{i}'}{\partial x_{j}\partial x_{j}}.
\end{multline}
The explicit spatial dependence can be removed by transforming coordinates. Transforming as
\begin{eqnarray}
x_{i} & = & \alpha\!\left(t\right)X_{i},\\
v_{i}'\!\left(\mathbf{x},t\right) & = & V_{i}\!\left(\mathbf{X},t\right),\\
p'\!\left(\mathbf{x},t\right) & = & P\!\left(\mathbf{X},t\right),
\end{eqnarray}
yields,
\begin{multline}
\rho_{0}\left(\frac{\partial V_{i}}{\partial t}+\frac{1}{\alpha}V_{j}\frac{\partial V_{i}}{\partial X_{j}}+\left(a-\frac{\dot{\alpha}}{\alpha}\right)X_{j}\frac{\partial V_{i}}{\partial X_{j}}+aV_{i}\right) =\\ 
-\frac{1}{\alpha}\frac{\partial}{\partial X_{j}}\left(\delta_{ij}P\right) + \frac{\mu\!\left(T\right)}{\alpha^{2}}\frac{\partial^{2}V_{i}}{\partial X_{j}\partial X_{j}}.\label{eq:moving_partway}
\end{multline}
Then, if the condition
\begin{equation}
a-\dot{\alpha}/\alpha=0 \label{eq:alpha_condition}
\end{equation}
is satisfied, the explicit spatial dependence is removed from the moving frame momentum equation, Eq.~(\ref{eq:moving_partway}) and it becomes,
\begin{multline}
\rho_{0}\left(\frac{\partial V_{i}}{\partial t}+\frac{1}{\alpha}V_{j}\frac{\partial V_{i}}{\partial X_{j}}+aV_{i}\right) = \\
- \frac{1}{\alpha}\frac{\partial}{\partial X_{j}}\left(\delta_{ij}P\right) +\frac{\mu\!\left(T\right)}{\alpha^{2}}\frac{\partial^{2}V_{i}}{\partial X_{j}\partial X_{j}}.\label{eq:moving_non_L}
\end{multline}
Together, the conditions Eq.~(\ref{eq:alpha_condition}) and Eq.~(\ref{eq:flow_condition}) say that $\ddot{\alpha}\!\left(t\right) = 0$. Consistent with this, define
\begin{eqnarray}
\alpha\!\left(t\right) &=& \left(L_{0}-2U_{b}t\right)/L_{0}=L\!\left(t\right)/L_{0},\\
L\!\left(t\right) &=& L_0 - 2 U_b t. \label{eq:L_def}
\end{eqnarray}
Then
\begin{equation}
a = \dot{L}/L,\label{eq:a}
\end{equation}
and the background flow ($\mathbf{v}_{0}$) is such that a cube of initial side length $L_{0}$, placed in the flow at $t_{0}=0$, will remain a cube and shrink in time at a constant rate while having a side length of $L\!\left(t\right)$. Using
$a$ from Eq.~(\ref{eq:a}) in Eq.~(\ref{eq:mean_density_exp}) gives the expected density dependence, Eq.~(\ref{eq:mean_density_solution}). Using the viscosity, density, and temperature solutions, Eqs.~(\ref{eq:mu},\ref{eq:mean_density_solution},\ref{eq:temperature_solution}), in the moving frame momentum equation, Eq.~(\ref{eq:moving_non_L}) gives the model equation Eq.~(\ref{eq:moving_momentum}).

\subsection{Scaled momentum equation}\label{sec:momentum_scaling}
The independent variables in Eq.~(\ref{eq:moving_momentum}) can be rescaled, and some time dependent coefficients eliminated. This is useful for simulations, and can be an aid in analysis. Using the scalings,
\begin{eqnarray}
V_{i} & = & \bar{L}^{\delta}\hat{V}_{i},\\
P & = & \bar{L}^{\eta}\hat{P},\\
\mbox{d}\hat{t} & = & \bar{L}^{\tau}\mbox{d}t,
\end{eqnarray}
in Eq.~(\ref{eq:moving_momentum}) gives,
\begin{multline}
\frac{\partial\hat{\mathbf{V}}}{\partial\hat{t}}+\bar{L}^{\delta-1-\tau}\hat{\mathbf{V}}\cdot \nabla \hat{\mathbf{V}} - 2 \bar{U}_b\bar{L}^{-\tau-1}\left(1+\delta\right)\hat{\mathbf{V}} = \\
-\bar{L}^{2+\eta-\delta-\tau}\nabla \hat{P}+ \frac{1}{\mbox{Re}_{0}}\bar{L}^{-2\beta-\tau+1} \nabla^2 \hat{\mathbf{V}}.\label{eq:momentum_scalings}
\end{multline}
The standard nondimensionalization has been used, so that $\mathrm{Re}_0=L_0 V_0/\nu_0$. Equation \ref{eq:momentum_scalings} has four independent powers of $\bar{L}$, and three undetermined
scaling factors, $\delta$, $\eta$, and $\tau$, so that the time
dependence can be eliminated from all but one term. One specific choice
takes $\delta=-1$ to eliminate the forcing term (with $U_{b}$ in
the coefficient), and then the time dependence of two other terms
can be eliminated. The choice where the forcing term and all time dependence but the viscosity's are eliminated has been discussed by Cambon et al.~\cite{cambon1992}). Choosing to eliminate the time dependence of all
but the forcing term, by selecting,
\begin{subequations} 
\label{eq:scalings}
\begin{align}
\tau & = 1-2\beta,\\
\delta & = 2-2\beta,\\
\eta & = 1-4\beta,
\end{align}
\end{subequations}
gives
\begin{equation}
\frac{\partial\hat{\mathbf{V}}}{\partial\hat{t}}+\hat{\mathbf{V}}\cdot \nabla \hat{\mathbf{V}} - 2 \bar{U}_b \bar{L}^{2\beta-2}\left(3-2\beta\right)\hat{\mathbf{V}} =
-\nabla \hat{P}+ \frac{1}{\mbox{Re}_{0}} \nabla^2 \hat{\mathbf{V}}.\label{eq:scaled_momentum}
\end{equation}

\subsection{$\beta = 1$ steady state energy} \label{sec:beta_1_steady}
A steady state solution ($\mathrm{d}E^T/\mathrm{d}t=0$) to the total energy equation, Eq.~(\ref{eq:integrated_energy}), when $\beta=1$, would mean
\begin{equation}
E^{T}_{\mathrm{steady}} = \frac{L_{0}}{4U_{b}}\epsilon_{\mathrm{steady}},\label{eq:equilibrium_condition}
\end{equation}
where $\epsilon_{\mathrm{steady}}=2\nu_{0}\int_{k_{\mathrm{min}}}^{\infty}\mbox{d}kk^{2}E\!\left(k\right)$
is the mean dissipation in steady state. When $\beta=1$, the scaled momentum
equation in the moving frame, Eq.~(\ref{eq:scaled_momentum}) is the usual Navier-Stokes equation with a time independent forcing. This equation has been studied in the context of a forcing scheme for isotropic fluid turbulence, where the term $2\bar{U}_{b}\hat{V}_{i}$ is added as an alternative to band-limited wavenumber space forcings~\cite{lundgren2003,rosales2005,carroll2013}. Numerical simulations by Rosales and Meneveau~\cite{rosales2005}
show that, in steady state, solutions
have a characteristic length scale, $l=u_{\mathrm{rms}}^{3}/\epsilon=0.19\mathcal{L}$,
where $\mathcal{L}$ is the domain size. Accounting for definitions and the scalings in Eqs.~(\ref{eq:scalings}), this relationship between $\epsilon_{\mathrm{steady}}$,$L_0$, and $E^T \propto u^2_{\mathrm{rms}}$ allows us to solve for $\epsilon_{\mathrm{steady}}$,
\begin{equation}
\epsilon_{\mathrm{steady}} = \left(\frac{2E^{T}_{\mathrm{steady}}}{3\rho_{0}\!\left(0\right)}\right)^{3/2}\frac{1}{0.19L_{0}}.\label{eq:epsilon_ss}
\end{equation}
Then Eqs.~(\ref{eq:equilibrium_condition}) and (\ref{eq:epsilon_ss}) can be solved for $E^T_{\mathrm{steady}}$, yielding Eq.~(\ref{eq:E_steady_Ub}) in section \ref{sec:analysis_beta1}.

\subsection{Temperature equation including bremsstrahlung} \label{sec:radiation}
Given here is a simple accounting of the effects of radiation without straying far from the present model. An optically thin plasma, with a single ion species of a single (time dependent) charge state $Z$ is assumed. The power density of electron bremsstrahlung emitted from an optically thin plasma, assuming $T_i = T_e = T$ and $n_e = Z n_i = Z n$, is
\begin{equation}
\mathrm{P}_{\mathrm{Br}}\left[W/m^3\right] = C_B \left(T\left[eV\right]\right)^{1/2} n^2 Z^3. \label{eq:bremsstrahlung}
\end{equation}
The bremsstrahlung constant is $C_B = 1.69101\times10^{-38} W \times m^3/\sqrt{eV}$. The internal energy equation for the isotropically compressed plasma, including the mechanical work and Bremsstrahlung terms only, and continuing to assume that the adiabatic index $\gamma = 5/3$, is
\begin{equation}
\frac{\partial}{\partial t} \left(\frac{3}{2} n_T k_B T \right) = - \frac{5}{2} n_T k_B T \left(3 \frac{\dot{L}}{L} \right) - C_B T^{1/2} n^{2} Z^{3}.\label{eq:internal_energy}
\end{equation}
Here $k_B$ is the Boltzmann constant, $\dot{L}$ can be found from Eq.~(\ref{eq:L_def}), and $n_T$ is the total number density, $n_T = n_i + n_e = \left(1+ Z \right) n$. Consistent with the spirit of the model described in Sec.~\ref{sec:model and energy} and the Appendix Sec.~\ref{sec:derivation}, the density is taken to be $n = n_0/\bar{L}^{3}$ (see Eq.~(\ref{eq:mean_density_solution})). Then, if the bremsstrahlung term in Eq.~(\ref{eq:internal_energy}) is ignored, the solution is $T = T_0/\bar{L}^2$, as in Eq.~(\ref{eq:temperature_solution}).

Rewriting Eq.~(\ref{eq:internal_energy}) as an equation for the normalized temperature, $\bar{T} = T/T_0$, as a function of the compression, while assuming that the charge state $Z$ is a function of temperature, gives
\begin{equation}
\frac{\partial \bar{T}}{\partial \bar{L}} = -2 \frac{\bar{T}}{\bar{L}} + 2 \frac{\tau_c}{\tau_{r,0}} \frac{2 Z^3}{1+Z} \bar{L}^{-3} \bar{T}^{1/2} - \frac{\bar{T}}{1+Z}\frac{\partial Z}{\partial \bar{T}}\frac{\partial \bar{T}}{\partial \bar{L}}. \label{eq:modified_T}
\end{equation}
The first term in Eq.~(\ref{eq:modified_T}) gives the mechanical heating (adiabatic heating when taken alone), while the second term represents bremsstrahlung cooling. The last term is associated with the energy needed to bring newly ionized electrons to the temperature $T$. If the charge state increases with temperature, and the temperature increases with compression (with decreasing $\bar{L}$) then it is a cooling term (acts to decrease the temperature). It should not, however, be taken as an accurate accounting of this energy. Our primary focus is comparing the radiation and adiabatic compression terms. The relative size of the radiation term is set by the compression time,
\begin{equation}
\tau_c = \frac{L_0}{2 U_b}
\end{equation}
and the initial radiation time,
\begin{equation}
\tau_{r}\left[s\right] = 3.01\times10^{-9} \frac{A_i T_{keV}^{1/2}}{\rho_{g/cc}}
\end{equation}
The ratio $\tau_c/\tau_{r}$ multiplied by the charge state coefficient $2 Z^3/\left(1+Z\right)$, gives the ratio of the bremsstrahlung cooling to the mechanical heating for any set of density, temperature, charge state, and ion mass number $A_i$. To solve for the temperature evolution as a function of compression, one evaluates the ratio at the initial temperature and density, as in Eq.~(\ref{eq:modified_T}), and solves that equation. For an arbitrary function $Z\left(T\right)$, the temperature will have some dependence on $L$, which can be used instead of the adiabatic relation $\bar{T} = \bar{L}^{-2}$ in the model described in Secs.~\ref{sec:model and energy},~\ref{sec:derivation}. Generally this will break the ability to reach a nicely scaled equation for the sake of simulation, Eq.~(\ref{eq:scaled_momentum}).

To give a simple example, consider the case where the charge state takes a simple power law relation with the temperature,
\begin{equation}
Z = Z_0 \bar{T}^{\phi}.
\end{equation}
Approximating $1+ Z \sim Z$, Eq.~(\ref{eq:modified_T}) can be reduced to
\begin{equation}
\left(1 + \phi \right) \frac{\partial \bar{T}}{\partial \bar{L}} = -2 \frac{\bar{T}}{\bar{L}} + 2 \frac{2 \tau_c}{\tau_{r,0}} \frac{\bar{T}^{2\phi + 1/2}}{\bar{L}^3}, \label{eq:general_phi_T}
\end{equation}
where the prefactor is due to the last term in Eq.~(\ref{eq:modified_T}) (the energy required to bring newly ionized electrons to temperature $\bar{T}$), and will result in an effective lower adiabatic index. However, this is not a radiation effect, and will be ignored for the discussion here. Equation \ref{eq:general_phi_T} can be solved analytically, and the solution takes a particularly simple form for $\phi = 1/4$, which has behavior that is qualitatively similar to the solutions for other $\phi$. When $\phi = 1/4$, and ignoring the prefactor on the derivative, the solution to Eq.~(\ref{eq:general_phi_T}) is
\begin{equation}
\bar{T}_{\phi = 1/4} = \frac{1}{\bar{L}^2} \mathrm{exp} \! \left[ \frac{2 \tau_c}{\tau_{r,0}} \left(1 - \bar{L}^{-2} \right) \right]
\end{equation}
For small initial compression time to radiation time ($\tau_c/\tau_{r,0} \ll 1$), the temperature tracks very closely with $1/\bar{L}^2$, up until the radiation becomes important -- $\bar{L} \sim \sqrt{2 \tau_c/\tau_{r}}$ for this $\phi=1/4$ case. Then, the model for turbulence behavior with ionization discussed in this work will be unmodified up until the point where the radiation becomes important. Provided that one starts the compression with a small initial $\tau_c/\tau_{r}$, this can hold for large compression ratios. Note that in this case, the temperature is no longer a state-function of compression, since it depends also on the compression rate. When the temperature tracks closely with $1/\bar{L}^2$, $\phi=1/4$ corresponds to the $\beta = 1.5$ case, for which simulation results are included in Figs.~\ref{fig:KEvsL_twoBetas} and~\ref{fig:vary_beta}. Radiation considerations are discussed further in Sec.~\ref{sec:discussion}.


\begin{thebibliography}{27}%
\makeatletter
\providecommand \@ifxundefined [1]{%
 \@ifx{#1\undefined}
}%
\providecommand \@ifnum [1]{%
 \ifnum #1\expandafter \@firstoftwo
 \else \expandafter \@secondoftwo
 \fi
}%
\providecommand \@ifx [1]{%
 \ifx #1\expandafter \@firstoftwo
 \else \expandafter \@secondoftwo
 \fi
}%
\providecommand \natexlab [1]{#1}%
\providecommand \enquote  [1]{``#1''}%
\providecommand \bibnamefont  [1]{#1}%
\providecommand \bibfnamefont [1]{#1}%
\providecommand \citenamefont [1]{#1}%
\providecommand \href@noop [0]{\@secondoftwo}%
\providecommand \href [0]{\begingroup \@sanitize@url \@href}%
\providecommand \@href[1]{\@@startlink{#1}\@@href}%
\providecommand \@@href[1]{\endgroup#1\@@endlink}%
\providecommand \@sanitize@url [0]{\catcode `\\12\catcode `\$12\catcode
  `\&12\catcode `\#12\catcode `\^12\catcode `\_12\catcode `\%12\relax}%
\providecommand \@@startlink[1]{}%
\providecommand \@@endlink[0]{}%
\providecommand \url  [0]{\begingroup\@sanitize@url \@url }%
\providecommand \@url [1]{\endgroup\@href {#1}{\urlprefix }}%
\providecommand \urlprefix  [0]{URL }%
\providecommand \Eprint [0]{\href }%
\providecommand \doibase [0]{http://dx.doi.org/}%
\providecommand \selectlanguage [0]{\@gobble}%
\providecommand \bibinfo  [0]{\@secondoftwo}%
\providecommand \bibfield  [0]{\@secondoftwo}%
\providecommand \translation [1]{[#1]}%
\providecommand \BibitemOpen [0]{}%
\providecommand \bibitemStop [0]{}%
\providecommand \bibitemNoStop [0]{.\EOS\space}%
\providecommand \EOS [0]{\spacefactor3000\relax}%
\providecommand \BibitemShut  [1]{\csname bibitem#1\endcsname}%
\let\auto@bib@innerbib\@empty
\bibitem [{\citenamefont {Davidovits}\ and\ \citenamefont
  {Fisch}(2016)}]{davidovits2016}%
  \BibitemOpen
  \bibfield  {author} {\bibinfo {author} {\bibfnamefont {S.}~\bibnamefont
  {Davidovits}}\ and\ \bibinfo {author} {\bibfnamefont {N.~J.}\ \bibnamefont
  {Fisch}},\ }\href {\doibase 10.1103/PhysRevLett.116.105004} {\bibfield
  {journal} {\bibinfo  {journal} {Phys. Rev. Lett.}\ }\textbf {\bibinfo
  {volume} {116}},\ \bibinfo {pages} {105004} (\bibinfo {year}
  {2016})}\BibitemShut {NoStop}%
\bibitem [{\citenamefont {Kroupp}\ \emph {et~al.}(2011)\citenamefont {Kroupp},
  \citenamefont {Osin}, \citenamefont {Starobinets}, \citenamefont {Fisher},
  \citenamefont {Bernshtam}, \citenamefont {Weingarten}, \citenamefont {Maron},
  \citenamefont {Uschmann}, \citenamefont {F\"orster}, \citenamefont {Fisher},
  \citenamefont {Cuneo}, \citenamefont {Deeney},\ and\ \citenamefont
  {Giuliani}}]{kroupp2011}%
  \BibitemOpen
  \bibfield  {author} {\bibinfo {author} {\bibfnamefont {E.}~\bibnamefont
  {Kroupp}}, \bibinfo {author} {\bibfnamefont {D.}~\bibnamefont {Osin}},
  \bibinfo {author} {\bibfnamefont {A.}~\bibnamefont {Starobinets}}, \bibinfo
  {author} {\bibfnamefont {V.}~\bibnamefont {Fisher}}, \bibinfo {author}
  {\bibfnamefont {V.}~\bibnamefont {Bernshtam}}, \bibinfo {author}
  {\bibfnamefont {L.}~\bibnamefont {Weingarten}}, \bibinfo {author}
  {\bibfnamefont {Y.}~\bibnamefont {Maron}}, \bibinfo {author} {\bibfnamefont
  {I.}~\bibnamefont {Uschmann}}, \bibinfo {author} {\bibfnamefont
  {E.}~\bibnamefont {F\"orster}}, \bibinfo {author} {\bibfnamefont
  {A.}~\bibnamefont {Fisher}}, \bibinfo {author} {\bibfnamefont {M.~E.}\
  \bibnamefont {Cuneo}}, \bibinfo {author} {\bibfnamefont {C.}~\bibnamefont
  {Deeney}}, \ and\ \bibinfo {author} {\bibfnamefont {J.~L.}\ \bibnamefont
  {Giuliani}},\ }\href {\doibase 10.1103/PhysRevLett.107.105001} {\bibfield
  {journal} {\bibinfo  {journal} {Phys. Rev. Lett.}\ }\textbf {\bibinfo
  {volume} {107}},\ \bibinfo {pages} {105001} (\bibinfo {year}
  {2011})}\BibitemShut {NoStop}%
\bibitem [{\citenamefont {Foord}\ \emph {et~al.}(1994)\citenamefont {Foord},
  \citenamefont {Maron}, \citenamefont {Davara}, \citenamefont {Gregorian},\
  and\ \citenamefont {Fisher}}]{foord1994}%
  \BibitemOpen
  \bibfield  {author} {\bibinfo {author} {\bibfnamefont {M.~E.}\ \bibnamefont
  {Foord}}, \bibinfo {author} {\bibfnamefont {Y.}~\bibnamefont {Maron}},
  \bibinfo {author} {\bibfnamefont {G.}~\bibnamefont {Davara}}, \bibinfo
  {author} {\bibfnamefont {L.}~\bibnamefont {Gregorian}}, \ and\ \bibinfo
  {author} {\bibfnamefont {A.}~\bibnamefont {Fisher}},\ }\href {\doibase
  10.1103/PhysRevLett.72.3827} {\bibfield  {journal} {\bibinfo  {journal}
  {Phys. Rev. Lett.}\ }\textbf {\bibinfo {volume} {72}},\ \bibinfo {pages}
  {3827} (\bibinfo {year} {1994})}\BibitemShut {NoStop}%
\bibitem [{\citenamefont {Kroupp}\ \emph
  {et~al.}(2007{\natexlab{a}})\citenamefont {Kroupp}, \citenamefont
  {Gregorian}, \citenamefont {Davara}, \citenamefont {Starobinets},
  \citenamefont {Stambulchik}, \citenamefont {Maron}, \citenamefont
  {Ralchenko},\ and\ \citenamefont {Alexiou}}]{kroupp2007}%
  \BibitemOpen
  \bibfield  {author} {\bibinfo {author} {\bibfnamefont {E.}~\bibnamefont
  {Kroupp}}, \bibinfo {author} {\bibfnamefont {L.}~\bibnamefont {Gregorian}},
  \bibinfo {author} {\bibfnamefont {G.}~\bibnamefont {Davara}}, \bibinfo
  {author} {\bibfnamefont {A.}~\bibnamefont {Starobinets}}, \bibinfo {author}
  {\bibfnamefont {E.}~\bibnamefont {Stambulchik}}, \bibinfo {author}
  {\bibfnamefont {Y.}~\bibnamefont {Maron}}, \bibinfo {author} {\bibfnamefont
  {Y.}~\bibnamefont {Ralchenko}}, \ and\ \bibinfo {author} {\bibfnamefont
  {S.}~\bibnamefont {Alexiou}},\ }\href@noop {} {\bibfield  {journal} {\bibinfo
   {journal} {AIP Conference Proceedings}\ }\textbf {\bibinfo {volume} {926}}
  (\bibinfo {year} {2007}{\natexlab{a}})}\BibitemShut {NoStop}%
\bibitem [{\citenamefont {Kroupp}\ \emph
  {et~al.}(2007{\natexlab{b}})\citenamefont {Kroupp}, \citenamefont {Osin},
  \citenamefont {Starobinets}, \citenamefont {Fisher}, \citenamefont
  {Bernshtam}, \citenamefont {Maron}, \citenamefont {Uschmann}, \citenamefont
  {F\"orster}, \citenamefont {Fisher},\ and\ \citenamefont
  {Deeney}}]{kroupp2007a}%
  \BibitemOpen
  \bibfield  {author} {\bibinfo {author} {\bibfnamefont {E.}~\bibnamefont
  {Kroupp}}, \bibinfo {author} {\bibfnamefont {D.}~\bibnamefont {Osin}},
  \bibinfo {author} {\bibfnamefont {A.}~\bibnamefont {Starobinets}}, \bibinfo
  {author} {\bibfnamefont {V.}~\bibnamefont {Fisher}}, \bibinfo {author}
  {\bibfnamefont {V.}~\bibnamefont {Bernshtam}}, \bibinfo {author}
  {\bibfnamefont {Y.}~\bibnamefont {Maron}}, \bibinfo {author} {\bibfnamefont
  {I.}~\bibnamefont {Uschmann}}, \bibinfo {author} {\bibfnamefont
  {E.}~\bibnamefont {F\"orster}}, \bibinfo {author} {\bibfnamefont
  {A.}~\bibnamefont {Fisher}}, \ and\ \bibinfo {author} {\bibfnamefont
  {C.}~\bibnamefont {Deeney}},\ }\href {\doibase 10.1103/PhysRevLett.98.115001}
  {\bibfield  {journal} {\bibinfo  {journal} {Phys. Rev. Lett.}\ }\textbf
  {\bibinfo {volume} {98}},\ \bibinfo {pages} {115001} (\bibinfo {year}
  {2007}{\natexlab{b}})}\BibitemShut {NoStop}%
\bibitem [{\citenamefont {Maron}\ \emph {et~al.}(2013)\citenamefont {Maron},
  \citenamefont {Starobinets}, \citenamefont {Fisher}, \citenamefont {Kroupp},
  \citenamefont {Osin}, \citenamefont {Fisher}, \citenamefont {Deeney},
  \citenamefont {Coverdale}, \citenamefont {Lepell}, \citenamefont {Yu},
  \citenamefont {Jennings}, \citenamefont {Cuneo}, \citenamefont {Herrmann},
  \citenamefont {Porter}, \citenamefont {Mehlhorn},\ and\ \citenamefont
  {Apruzese}}]{maron2013}%
  \BibitemOpen
  \bibfield  {author} {\bibinfo {author} {\bibfnamefont {Y.}~\bibnamefont
  {Maron}}, \bibinfo {author} {\bibfnamefont {A.}~\bibnamefont {Starobinets}},
  \bibinfo {author} {\bibfnamefont {V.~I.}\ \bibnamefont {Fisher}}, \bibinfo
  {author} {\bibfnamefont {E.}~\bibnamefont {Kroupp}}, \bibinfo {author}
  {\bibfnamefont {D.}~\bibnamefont {Osin}}, \bibinfo {author} {\bibfnamefont
  {A.}~\bibnamefont {Fisher}}, \bibinfo {author} {\bibfnamefont
  {C.}~\bibnamefont {Deeney}}, \bibinfo {author} {\bibfnamefont {C.~A.}\
  \bibnamefont {Coverdale}}, \bibinfo {author} {\bibfnamefont {P.~D.}\
  \bibnamefont {Lepell}}, \bibinfo {author} {\bibfnamefont {E.~P.}\
  \bibnamefont {Yu}}, \bibinfo {author} {\bibfnamefont {C.}~\bibnamefont
  {Jennings}}, \bibinfo {author} {\bibfnamefont {M.~E.}\ \bibnamefont {Cuneo}},
  \bibinfo {author} {\bibfnamefont {M.~C.}\ \bibnamefont {Herrmann}}, \bibinfo
  {author} {\bibfnamefont {J.~L.}\ \bibnamefont {Porter}}, \bibinfo {author}
  {\bibfnamefont {T.~A.}\ \bibnamefont {Mehlhorn}}, \ and\ \bibinfo {author}
  {\bibfnamefont {J.~P.}\ \bibnamefont {Apruzese}},\ }\href {\doibase
  10.1103/PhysRevLett.111.035001} {\bibfield  {journal} {\bibinfo  {journal}
  {Phys. Rev. Lett.}\ }\textbf {\bibinfo {volume} {111}},\ \bibinfo {pages}
  {035001} (\bibinfo {year} {2013})}\BibitemShut {NoStop}%
\bibitem [{\citenamefont {Thomas}\ and\ \citenamefont
  {Kares}(2012)}]{thomas2012}%
  \BibitemOpen
  \bibfield  {author} {\bibinfo {author} {\bibfnamefont {V.~A.}\ \bibnamefont
  {Thomas}}\ and\ \bibinfo {author} {\bibfnamefont {R.~J.}\ \bibnamefont
  {Kares}},\ }\href {\doibase 10.1103/PhysRevLett.109.075004} {\bibfield
  {journal} {\bibinfo  {journal} {Phys. Rev. Lett.}\ }\textbf {\bibinfo
  {volume} {109}},\ \bibinfo {pages} {075004} (\bibinfo {year}
  {2012})}\BibitemShut {NoStop}%
\bibitem [{\citenamefont {Weber}\ \emph {et~al.}(2014)\citenamefont {Weber},
  \citenamefont {Clark}, \citenamefont {Cook}, \citenamefont {Busby},\ and\
  \citenamefont {Robey}}]{weber2014}%
  \BibitemOpen
  \bibfield  {author} {\bibinfo {author} {\bibfnamefont {C.~R.}\ \bibnamefont
  {Weber}}, \bibinfo {author} {\bibfnamefont {D.~S.}\ \bibnamefont {Clark}},
  \bibinfo {author} {\bibfnamefont {A.~W.}\ \bibnamefont {Cook}}, \bibinfo
  {author} {\bibfnamefont {L.~E.}\ \bibnamefont {Busby}}, \ and\ \bibinfo
  {author} {\bibfnamefont {H.~F.}\ \bibnamefont {Robey}},\ }\href {\doibase
  10.1103/PhysRevE.89.053106} {\bibfield  {journal} {\bibinfo  {journal} {Phys.
  Rev. E}\ }\textbf {\bibinfo {volume} {89}},\ \bibinfo {pages} {053106}
  (\bibinfo {year} {2014})}\BibitemShut {NoStop}%
\bibitem [{\citenamefont {Wu}\ \emph {et~al.}(1985)\citenamefont {Wu},
  \citenamefont {Ferziger},\ and\ \citenamefont {Chapman}}]{wu1985}%
  \BibitemOpen
  \bibfield  {author} {\bibinfo {author} {\bibfnamefont {C.-T.}\ \bibnamefont
  {Wu}}, \bibinfo {author} {\bibfnamefont {J.~H.}\ \bibnamefont {Ferziger}}, \
  and\ \bibinfo {author} {\bibfnamefont {D.~R.}\ \bibnamefont {Chapman}},\
  }\href@noop {} {\bibfield  {journal} {\bibinfo  {journal} {Stanford
  University, Department of Mechanical Engineering Report No. TF-21}\ }
  (\bibinfo {year} {1985})}\BibitemShut {NoStop}%
\bibitem [{\citenamefont {Coleman}\ and\ \citenamefont
  {Mansour}(1991)}]{coleman1991}%
  \BibitemOpen
  \bibfield  {author} {\bibinfo {author} {\bibfnamefont {G.~N.}\ \bibnamefont
  {Coleman}}\ and\ \bibinfo {author} {\bibfnamefont {N.~N.}\ \bibnamefont
  {Mansour}},\ }\href@noop {} {\bibfield  {journal} {\bibinfo  {journal}
  {Physics of Fluids A}\ }\textbf {\bibinfo {volume} {3}} (\bibinfo {year}
  {1991})}\BibitemShut {NoStop}%
\bibitem [{\citenamefont {{Blaisdell}}(1991)}]{blaisdell1991}%
  \BibitemOpen
  \bibfield  {author} {\bibinfo {author} {\bibfnamefont {G.~A.}\ \bibnamefont
  {{Blaisdell}}},\ }\emph {\bibinfo {title} {{Numerical simulation of
  compressible homogeneous turbulence}}},\ \href@noop {} {Ph.D. thesis},\
  \bibinfo  {school} {Stanford Univ., CA.} (\bibinfo {year} {1991})\BibitemShut
  {NoStop}%
\bibitem [{\citenamefont {Cambon}\ \emph {et~al.}(1992)\citenamefont {Cambon},
  \citenamefont {Mao},\ and\ \citenamefont {Jeandel}}]{cambon1992}%
  \BibitemOpen
  \bibfield  {author} {\bibinfo {author} {\bibfnamefont {C.}~\bibnamefont
  {Cambon}}, \bibinfo {author} {\bibfnamefont {Y.}~\bibnamefont {Mao}}, \ and\
  \bibinfo {author} {\bibfnamefont {D.}~\bibnamefont {Jeandel}},\ }\href
  {http://adsabs.harvard.edu/abs/1992EJMF...11..683C} {\bibfield  {journal}
  {\bibinfo  {journal} {European Journal of Mechanics - B/Fluids}\ }\textbf
  {\bibinfo {volume} {11}},\ \bibinfo {pages} {683} (\bibinfo {year}
  {1992})}\BibitemShut {NoStop}%
\bibitem [{\citenamefont {Hamlington}\ and\ \citenamefont
  {Ihme}(2014)}]{hamlington2014}%
  \BibitemOpen
  \bibfield  {author} {\bibinfo {author} {\bibfnamefont {P.}~\bibnamefont
  {Hamlington}}\ and\ \bibinfo {author} {\bibfnamefont {M.}~\bibnamefont
  {Ihme}},\ }\href {\doibase 10.1007/s10494-014-9535-7} {\bibfield  {journal}
  {\bibinfo  {journal} {Flow, Turbulence and Combustion}\ }\textbf {\bibinfo
  {volume} {93}},\ \bibinfo {pages} {93} (\bibinfo {year} {2014})}\BibitemShut
  {NoStop}%
\bibitem [{\citenamefont {McComb}(1990)}]{mccomb1990}%
  \BibitemOpen
  \bibfield  {author} {\bibinfo {author} {\bibfnamefont {W.~D.}\ \bibnamefont
  {McComb}},\ }\href@noop {} {\emph {\bibinfo {title} {The physics of fluid
  turbulence}}}\ (\bibinfo  {publisher} {Oxford University Press},\ \bibinfo
  {address} {Oxford: New York: Clarendon Press},\ \bibinfo {year}
  {1990})\BibitemShut {NoStop}%
\bibitem [{\citenamefont {Durbin}\ and\ \citenamefont
  {Reif}(2010)}]{durbin2010}%
  \BibitemOpen
  \bibfield  {author} {\bibinfo {author} {\bibfnamefont {P.}~\bibnamefont
  {Durbin}}\ and\ \bibinfo {author} {\bibfnamefont {B.}~\bibnamefont {Reif}},\
  }\href@noop {} {\emph {\bibinfo {title} {Statistical Theory and Modeling for
  Turbulent Flows}}}\ (\bibinfo  {publisher} {Wiley},\ \bibinfo {year}
  {2010})\BibitemShut {NoStop}%
\bibitem [{\citenamefont {Savill}(1987)}]{savill1987}%
  \BibitemOpen
  \bibfield  {author} {\bibinfo {author} {\bibfnamefont {A.~M.}\ \bibnamefont
  {Savill}},\ }\href {\doibase 10.1146/annurev.fl.19.010187.002531} {\bibfield
  {journal} {\bibinfo  {journal} {Annual Review of Fluid Mechanics}\ }\textbf
  {\bibinfo {volume} {19}},\ \bibinfo {pages} {531} (\bibinfo {year} {1987})},\
  \Eprint
  {http://arxiv.org/abs/http://dx.doi.org/10.1146/annurev.fl.19.010187.002531}
  {http://dx.doi.org/10.1146/annurev.fl.19.010187.002531} \BibitemShut
  {NoStop}%
\bibitem [{\citenamefont {Hunt}\ and\ \citenamefont
  {Carruthers}(1990)}]{hunt1990}%
  \BibitemOpen
  \bibfield  {author} {\bibinfo {author} {\bibfnamefont {J.~C.~R.}\
  \bibnamefont {Hunt}}\ and\ \bibinfo {author} {\bibfnamefont {D.~J.}\
  \bibnamefont {Carruthers}},\ }\href {\doibase 10.1017/S0022112090002075}
  {\bibfield  {journal} {\bibinfo  {journal} {Journal of Fluid Mechanics}\
  }\textbf {\bibinfo {volume} {212}},\ \bibinfo {pages} {497} (\bibinfo {year}
  {1990})}\BibitemShut {NoStop}%
\bibitem [{\citenamefont {Rosales}\ and\ \citenamefont
  {Meneveau}(2005)}]{rosales2005}%
  \BibitemOpen
  \bibfield  {author} {\bibinfo {author} {\bibfnamefont {C.}~\bibnamefont
  {Rosales}}\ and\ \bibinfo {author} {\bibfnamefont {C.}~\bibnamefont
  {Meneveau}},\ }\href {\doibase http://dx.doi.org/10.1063/1.2047568}
  {\bibfield  {journal} {\bibinfo  {journal} {Physics of Fluids}\ }\textbf
  {\bibinfo {volume} {17}},\ \bibinfo {eid} {095106} (\bibinfo {year}
  {2005})}\BibitemShut {NoStop}%
\bibitem [{\citenamefont {Coleman}\ and\ \citenamefont
  {Mansour}(1993)}]{coleman1993}%
  \BibitemOpen
  \bibfield  {author} {\bibinfo {author} {\bibfnamefont {G.}~\bibnamefont
  {Coleman}}\ and\ \bibinfo {author} {\bibfnamefont {N.}~\bibnamefont
  {Mansour}},\ }in\ \href {\doibase 10.1007/978-3-642-77674-8_19} {\emph
  {\bibinfo {booktitle} {Turbulent Shear Flows 8}}},\ \bibinfo {editor} {edited
  by\ \bibinfo {editor} {\bibfnamefont {F.}~\bibnamefont {Durst}}, \bibinfo
  {editor} {\bibfnamefont {R.}~\bibnamefont {Friedrich}}, \bibinfo {editor}
  {\bibfnamefont {B.}~\bibnamefont {Launder}}, \bibinfo {editor} {\bibfnamefont
  {F.}~\bibnamefont {Schmidt}}, \bibinfo {editor} {\bibfnamefont
  {U.}~\bibnamefont {Schumann}}, \ and\ \bibinfo {editor} {\bibfnamefont
  {J.}~\bibnamefont {Whitelaw}}}\ (\bibinfo  {publisher} {Springer Berlin
  Heidelberg},\ \bibinfo {year} {1993})\ pp.\ \bibinfo {pages}
  {269--282}\BibitemShut {NoStop}%
\bibitem [{ded()}]{dedalus}%
  \BibitemOpen
  \href@noop {} {\bibinfo  {journal} {http://dedalus-project.org/}\
  }\BibitemShut {NoStop}%
\bibitem [{\citenamefont {Lundgren}(2003)}]{lundgren2003}%
  \BibitemOpen
\bibfield  {journal} {  }\bibfield  {author} {\bibinfo {author} {\bibfnamefont
  {T.~S.}\ \bibnamefont {Lundgren}},\ }in\ \href@noop {} {\emph {\bibinfo
  {booktitle} {Annual Research Briefs}}}\ (\bibinfo  {publisher} {Center for
  Turbulence Research, Stanford},\ \bibinfo {year} {2003})\ pp.\ \bibinfo
  {pages} {461--473}\BibitemShut {NoStop}%
\bibitem [{\citenamefont {Schekochihin}\ \emph {et~al.}(2007)\citenamefont
  {Schekochihin}, \citenamefont {Iskakov}, \citenamefont {Cowley},
  \citenamefont {McWilliams}, \citenamefont {Proctor},\ and\ \citenamefont
  {Yousef}}]{schekochihin2007}%
  \BibitemOpen
  \bibfield  {author} {\bibinfo {author} {\bibfnamefont {A.~A.}\ \bibnamefont
  {Schekochihin}}, \bibinfo {author} {\bibfnamefont {A.~B.}\ \bibnamefont
  {Iskakov}}, \bibinfo {author} {\bibfnamefont {S.~C.}\ \bibnamefont {Cowley}},
  \bibinfo {author} {\bibfnamefont {J.~C.}\ \bibnamefont {McWilliams}},
  \bibinfo {author} {\bibfnamefont {M.~R.~E.}\ \bibnamefont {Proctor}}, \ and\
  \bibinfo {author} {\bibfnamefont {T.~A.}\ \bibnamefont {Yousef}},\ }\href
  {http://stacks.iop.org/1367-2630/9/i=8/a=300} {\bibfield  {journal} {\bibinfo
   {journal} {New Journal of Physics}\ }\textbf {\bibinfo {volume} {9}},\
  \bibinfo {pages} {300} (\bibinfo {year} {2007})}\BibitemShut {NoStop}%
\bibitem [{\citenamefont {Iskakov}\ \emph {et~al.}(2007)\citenamefont
  {Iskakov}, \citenamefont {Schekochihin}, \citenamefont {Cowley},
  \citenamefont {McWilliams},\ and\ \citenamefont {Proctor}}]{iskakov2007}%
  \BibitemOpen
  \bibfield  {author} {\bibinfo {author} {\bibfnamefont {A.~B.}\ \bibnamefont
  {Iskakov}}, \bibinfo {author} {\bibfnamefont {A.~A.}\ \bibnamefont
  {Schekochihin}}, \bibinfo {author} {\bibfnamefont {S.~C.}\ \bibnamefont
  {Cowley}}, \bibinfo {author} {\bibfnamefont {J.~C.}\ \bibnamefont
  {McWilliams}}, \ and\ \bibinfo {author} {\bibfnamefont {M.~R.~E.}\
  \bibnamefont {Proctor}},\ }\href {\doibase 10.1103/PhysRevLett.98.208501}
  {\bibfield  {journal} {\bibinfo  {journal} {Phys. Rev. Lett.}\ }\textbf
  {\bibinfo {volume} {98}},\ \bibinfo {pages} {208501} (\bibinfo {year}
  {2007})}\BibitemShut {NoStop}%
\bibitem [{\citenamefont {Brandenburg}(2014)}]{brandenburg2014}%
  \BibitemOpen
  \bibfield  {author} {\bibinfo {author} {\bibfnamefont {A.}~\bibnamefont
  {Brandenburg}},\ }\href {http://stacks.iop.org/0004-637X/791/i=1/a=12}
  {\bibfield  {journal} {\bibinfo  {journal} {The Astrophysical Journal}\
  }\textbf {\bibinfo {volume} {791}},\ \bibinfo {pages} {12} (\bibinfo {year}
  {2014})}\BibitemShut {NoStop}%
\bibitem [{\citenamefont {Lindl}(1995)}]{lindl1995}%
  \BibitemOpen
  \bibfield  {author} {\bibinfo {author} {\bibfnamefont {J.}~\bibnamefont
  {Lindl}},\ }\href@noop {} {\bibfield  {journal} {\bibinfo  {journal} {Physics
  of Plasmas}\ }\textbf {\bibinfo {volume} {2}} (\bibinfo {year}
  {1995})}\BibitemShut {NoStop}%
\bibitem [{\citenamefont {{Braginskii}}(1965)}]{braginskii1965}%
  \BibitemOpen
  \bibfield  {author} {\bibinfo {author} {\bibfnamefont {S.~I.}\ \bibnamefont
  {{Braginskii}}},\ }\href@noop {} {\bibfield  {journal} {\bibinfo  {journal}
  {Reviews of Plasma Physics}\ }\textbf {\bibinfo {volume} {1}},\ \bibinfo
  {pages} {205} (\bibinfo {year} {1965})}\BibitemShut {NoStop}%
\bibitem [{\citenamefont {Carroll}\ and\ \citenamefont
  {Blanquart}(2013)}]{carroll2013}%
  \BibitemOpen
  \bibfield  {author} {\bibinfo {author} {\bibfnamefont {P.~L.}\ \bibnamefont
  {Carroll}}\ and\ \bibinfo {author} {\bibfnamefont {G.}~\bibnamefont
  {Blanquart}},\ }\href {\doibase http://dx.doi.org/10.1063/1.4826315}
  {\bibfield  {journal} {\bibinfo  {journal} {Physics of Fluids}\ }\textbf
  {\bibinfo {volume} {25}},\ \bibinfo {eid} {105114} (\bibinfo {year}
  {2013})}\BibitemShut {NoStop}%
\end{thebibliography}
\end{document}